\documentclass[aps,reprint,nofootinbib,superscriptaddress]{revtex4-1}

\usepackage[english]{babel}
\usepackage[utf8x]{inputenc}
\usepackage[T1]{fontenc}

\usepackage{float}
\usepackage{amsmath}
\usepackage{amssymb}
\usepackage{hyperref}
\usepackage{cleveref}
\usepackage{graphicx}
\usepackage{glossaries}

\usepackage{physics}
\usepackage{siunitx}

\newcommand{\delay}[1]{\mathbf{D}_{#1}}
\newcommand{\ddelay}[1]{\dot{\mathbf{D}}_{#1}}
\newcommand{\tdidelay}[1]{\bar{\mathbf{D}}_{#1}}
\newcommand{\dtdidelay}[1]{\dot{\bar{\mathbf{D}}}_{#1}}

\newcommand{\psd}[1]{\opbraces{S_{#1}}}

\renewcommand{\qc}{\,\text{,}}
\newcommand{\qs}{\,\text{.}}

\newcommand{\isc}[0]{\text{isi}}
\newcommand{\rfi}[0]{\text{rfi}}
\newcommand{\tmi}[0]{\text{tmi}}
\newcommand{\prn}[0]{\text{PRN}}

\newacronym{lisa}{LISA}{Laser Interferometer Space Antenna}
\newacronym{mosa}{MOSA}{movable optical sub-assembly}
\newacronym{gw}{GW}{gravitational wave}
\newacronym{esa}{ESA}{European Space Agency}
\newacronym{gr}{GR}{general relativity}
\newacronym{tdi}{TDI}{time-delay interferometry}
\newacronym{prn}{PRN}{pseudo-random noise}
\newacronym{uso}{USO}{ultra-stable oscillator}
\newacronym{inrep}{INReP}{initial noise reduction pipeline}

\newacronym{tcb}{TCB}{Barycentric Coordinate Time}
\newacronym{bcrs}{BCRS}{Barycentric Celestial Reference System}
\newacronym{ltt}{LTT}{light travel time}
\newacronym{mpr}{MPR}{measured pseudorange}

\newacronym[longplural={power spectral densities}]{psd}{PSD}{power spectral density}
\newacronym[longplural={amplitude spectral densities}]{asd}{ASD}{amplitude spectral density}

\newacronym{rms}{RMS}{root mean square}

\begin{document}

\title{Time-delay interferometry without clock synchronisation}

\author{Olaf Hartwig}
\email{olaf.hartwig@obspm.fr}
\affiliation{SYRTE, Observatoire  de  Paris,  Universit\'e  PSL, CNRS,  Sorbonne  Universit\'e,  LNE,  61 avenue de l’Observatoire 75014  Paris,  France}
\affiliation{Max-Planck-Institut für Gravitationsphysik (Albert-Einstein-Institut),
Callinstraße 38, 30167 Hannover, Germany}

\author{Jean-Baptiste Bayle}
\affiliation{Jet Propulsion Laboratory, California Institute of Technology, 4800 Oak Grove Drive, Pasadena, CA 91109, USA}

\author{Martin Staab}
\affiliation{Max-Planck-Institut für Gravitationsphysik (Albert-Einstein-Institut),
Callinstraße 38, 30167 Hannover, Germany}

\author{Aur\'elien Hees}
\affiliation{SYRTE, Observatoire  de  Paris,  Universit\'e  PSL, CNRS,  Sorbonne  Universit\'e,  LNE,  61 avenue de l’Observatoire 75014  Paris,  France}

\author{Marc Lilley}
\affiliation{SYRTE, Observatoire  de  Paris,  Universit\'e  PSL, CNRS,  Sorbonne  Universit\'e,  LNE,  61 avenue de l’Observatoire 75014  Paris,  France}

\author{Peter Wolf}
\affiliation{SYRTE, Observatoire  de  Paris,  Universit\'e  PSL, CNRS,  Sorbonne  Universit\'e,  LNE,  61 avenue de l’Observatoire 75014  Paris,  France}

\date{\today}

\pacs{95.75.-z}
\keywords{lisa,time-delay interferometry,tdi,noise reduction,inrep,simulation}

\begin{abstract}
\Gls{tdi} is a data processing technique for \acrshort{lisa} designed to suppress the otherwise overwhelming laser noise by several orders of magnitude. It is widely believed that \gls{tdi} can only be applied once all phase or frequency measurements from each spacecraft have been synchronized to a common time frame. We demonstrate analytically, using as an example the commonly-used Michelson combination $X$, that \gls{tdi} can be computed using the raw, unsynchronized data, thereby avoiding the need for an initial synchronization processing step and significantly simplifying the initial noise reduction pipeline. Furthermore, the raw data is free of any potential artifacts introduced by clock synchronization and reference frame transformation algorithms, which allows to operate directly on the \si{\mega\Hz} beatnotes. As a consequence, in-band clock noise is directly suppressed as part of \gls{tdi}, in contrast to the approach previously proposed in the literature (in which large trends in the beatnotes are removed before the main laser-noise reduction step, and clock noise is suppressed in an extra processing step). We validate our algorithm with full-scale numerical simulations that use \texttt{LISA Instrument} and \texttt{PyTDI} and show that we reach the same performance levels as the previously proposed methods, ultimately limited by the clock sideband stability.

\end{abstract}

\maketitle
\glsresetall

\section{Introduction}

The first gravitational-wave detection of a binary black-hole merger~\cite{LIGOScientific:2016aoc} opened the era of gravitational astronomy, and its many promises. Gravitational waves carry crucial information about dense astrophysical systems, and their detection will bring answers to many questions ranging from the formation of the most massive objects to tests of the general theory of relativity. As of today, many more gravitational-wave events have been confirmed by ground-based detectors~\cite{LIGOScientific:2018mvr,LIGOScientific:2020ibl,LIGOScientific:2021djp}. Unfortunately these ground-based detectors are limited to signals above tens of \si{\hertz}, leaving out a large part of the gravitational-wave spectrum. Space-borne detectors, on the contrary, will be able to measure \si{\milli\hertz} gravitational-wave signals originating from various types of sources, such as supermassive black-hole binaries, Galactic black-hole or neutron star binaries, or cosmological defects from the early Universe.

The \gls{lisa} is one such mission led by the \gls{esa}, with an expected launch date in 2034~\cite{Audley:2017drz}. Three spacecraft in a nearly-equilateral triangular formation will trail the Earth on its heliocentric orbit. Each spacecraft will host two free-falling test masses and monitor their relative motion using laser interferometry. We expect that various noises of instrumental origin will enter the measurements at levels that violate the requirements. To reduce these noises to acceptable levels, a number of data processing algorithms are being developed as part of the multi-step \gls{inrep}. As an example, each spacecraft hosts an onboard clock that drives the sampling of the measurements. Because these clocks are not actively synchronized, they will jitter and drift with respect to each other and to a common time scale \cite{pireaux:2007sh}. One step of \gls{inrep} is to estimate these drifts and ultimately synchronize all measurements to a common time scale to prepare for the science analyses. Another crucial part of \gls{inrep} is a data processing algorithm called \gls{tdi}, which combines time-shifted measurements to reduce the otherwise overwhelming laser frequency noise by more than 8 orders of magnitude~\cite{Giampieri:1996aa,Armstrong:1999hp}.

The \textit{standard} procedure for carrying out \gls{tdi} relies on measurements that have previously all been synchronized to a common timescale \cite{Tinto:2020fcc}. The synchronization is achieved using auxiliary data measured on board the spacecraft~\cite{Heinzel_2011}, as well as the on-ground spacecraft tracking data. Given the uncertainties in those auxiliary measurements, some additional processing steps are required in \gls{inrep}, like the removal of large phase ramps and clock-noise correction~\cite{Hartwig:2020tdu}.

In this paper, we demonstrate that \gls{tdi} with full laser noise reduction can be achieved \textit{without} preliminary clock synchronization and \textit{without} relying on ground data from spacecraft tracking. The underlying reason for this is that \gls{tdi} does not fundamentally require the choice of any reference frame, as already pointed out in~\cite{Muratore:2020mdf}. Additional processing steps (removal of large phase ramps and clock-noise correction) in \gls{inrep} are avoided, significantly simplifying the procedure.

We present here an in-depth study of this alternative algorithm, and show algebraically and with numerical simulations that, indeed, \gls{tdi} reduces laser noise to required levels when used with raw data.

This paper is organized as follows. In \cref{sec:theory}, we review the standard \gls{tdi} algorithms and then show analytically that an alternative pipeline also produces laser noise-free data. In \cref{sec:sim_model}, we describe the instrumental setup we used in our numerical simulations; the results are presented and discussed in \cref{sec:sim_results}. We conclude in \cref{sec:conclusion}.

\section{Theoretical description}
\label{sec:theory}

\begin{figure}
    \centering
    \includegraphics[width=\columnwidth]{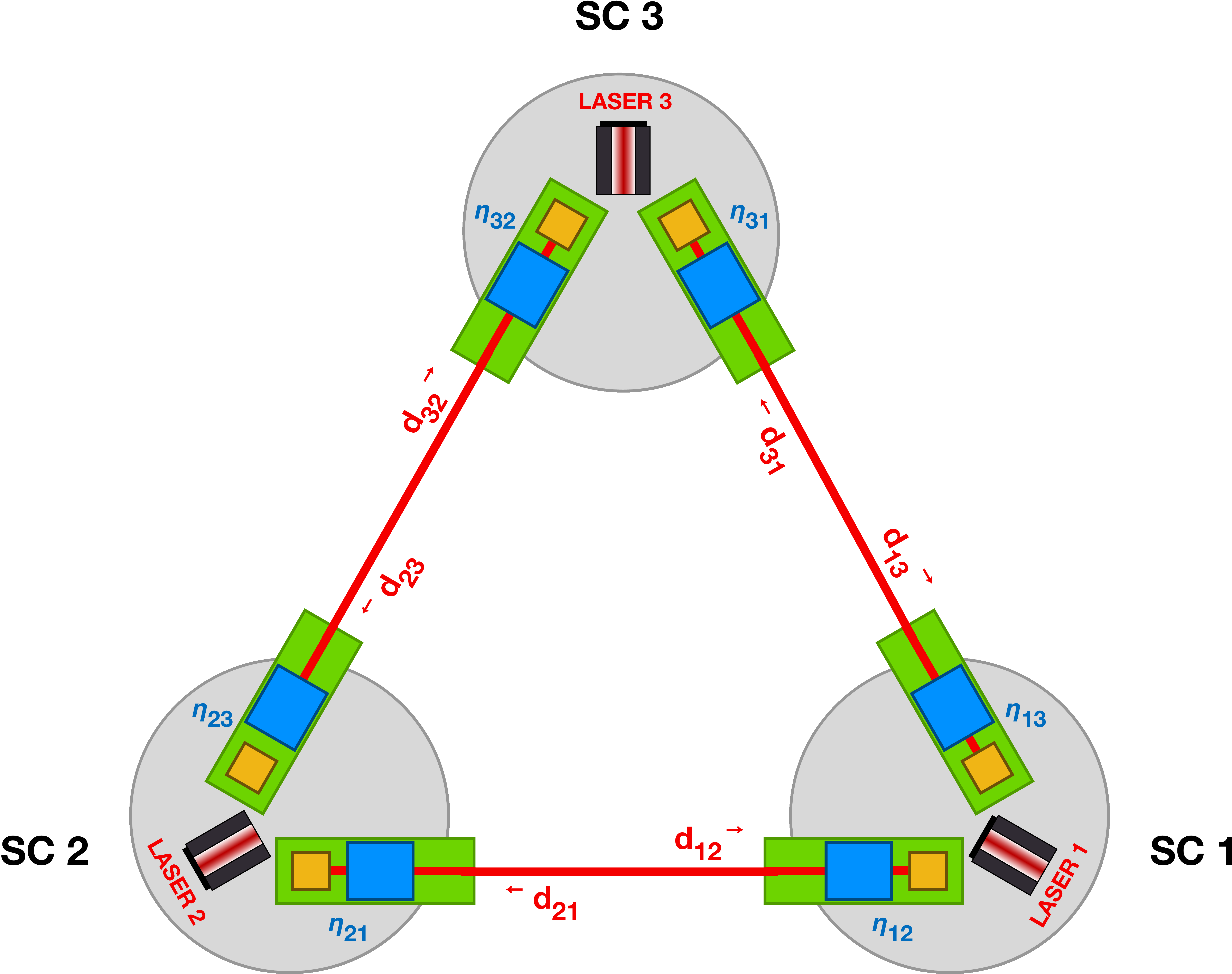}
    \caption{Simplified model of the instrument, as seen when looking down at the solar panels, with a total of 3 laser sources.}
    \label{fig:indexing}
\end{figure}

In this section, we consider a simplified model of our instrument, illustrated in \cref{fig:indexing}. The spacecraft are labeled 1, 2 and 3 clockwise when looking down at their solar panels. Each spacecraft~$i$ hosts a single laser source\footnote{In reality, each spacecraft carries 2 laser sources. However, auxiliary interferometers measure their local phase difference and it is thus possible to combine the measurements to reduce the problem to a total of three lasers~\cite{Otto:2012dk}. We consider this more complex model for the numerical simulations described in \cref{sec:sim_model}.} locked to a cavity\footnote{In reality, laser sources are locked to each other, and ultimately to a single resonant cavity. For simplicity, we ignore this in our description and assume that all lasers are independent.}.

We use three different time scales to describe the \gls{lisa} measurements and the associated data processing. The \gls{tcb}, denoted by $t$, is the coordinate time associated with the \gls{bcrs}. It is used to describe Solar System-scale phenomena, such as the spacecraft orbits~\cite{soffel:2003bd}. The three spacecraft proper times, denoted by $\tau_i$, are used to describe the physics inside each of these spacecraft. They depend on the spacecraft orbits because of relativistic effects. Their evolution with respect to the \gls{tcb} includes a drift of about \SI{0.4}{\second\per yr} and annual oscillations of the order of \SI{0.8}{\milli\second}~\cite{pireaux:2007sh}. Lastly, the three onboard clock times are denoted by $\hat\tau_i$. They define the times measured by the actual clocks onboard the spacecraft, and are the only timescales directly accessible from the measurements. In particular, they are used for timestamping and to generate the laser sidebands and the \gls{prn} code used for ranging~\cite{Heinzel_2011}. The onboard clock times differ from the spacecraft proper times by jitters and drifts of the clocks, both of instrumental origin. The transformation between $t$ and $\hat\tau_i$ will be denoted\footnote{The notation $x^{y}(\tau)$ indicates a quantity $x$ expressed in the timeframe $y$. E.g., $\hat\tau_i^t(\tau)$ has its argument (muted variable) $\tau$ expressed in the \gls{tcb} time $t$.} by 
\begin{equation}
    \hat\tau_i^t(\tau) = \tau + \delta \hat \tau_i^{t}(\tau)
    \qc
    \label{eq:clock_dev}
\end{equation}
and includes the relativistic effects relating $t$ and $\tau_i$, as well as the instrumental effects relating $\tau_i$ and $\hat\tau_i$.
The inverse operation, giving the \gls{tcb} time as a function of the clock time of spacecraft $i$, is given as
\begin{equation}
    t^{\hat\tau_i}(\tau) = \tau + \delta t^{\hat \tau_i}(\tau)
    \qs
    \label{eq:clock_dev_inv}
\end{equation}

\subsection{Review of standard TDI}
\label{sub:standard-tdi}

\begin{figure}
    \centering
    \includegraphics[width=\columnwidth]{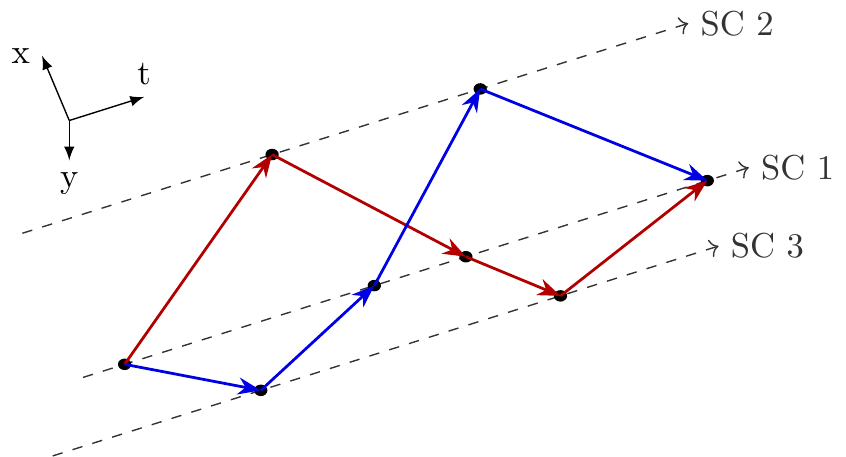}
    \caption{\Gls{tcb} spacetime diagram illustrating the geometrical interpretation of \gls{tdi}, as described in \cite{Vallisneri:2005ji,Muratore:2020mdf}. The dashed lines represent the world lines of the 3 spacecraft, while the blue and red lines display the paths of the two synthesized laser beams making up the first-generation Michelson combination $X_1$, before they finally interfere onboard spacecraft 1. The black dots mark the events of reception and emission of the laser beams.}
    \label{fig:spacetime}
\end{figure}

\Gls{tdi} is designed to strongly suppress the overwhelming laser frequency noise terms appearing in the \gls{lisa} measurements. \Gls{tdi} uses the fact that the same laser noise terms appear in different measurements, evaluated at different times, such that we can form linear combinations of these measurements in which all noise terms cancel. The appropriate time-shifts to apply relate the reception and emission events of the laser light exchanged between the spacecraft, as these are the events at which the laser noise enters our measurements. This idea was developed in~\cite{Vallisneri:2005ji} into \textit{geometric} \gls{tdi}, where \gls{tdi} combinations are interpreted as the interference of two (or more) virtual laser beams, which are aligned in time such that laser noise is cancelled at all reception and emission events. This is illustrated in the spacetime diagram in \cref{fig:spacetime}.

We call \textit{standard} \gls{tdi} the realization of \gls{tdi} that one can find in most of the literature (see, e.g.,~\cite{Tinto:2020fcc}, and references therein). It assumes that all measurements are synchronized to some common time frame, which is also the baseline implementation as part of the \gls{lisa} ground segment. This common time frame is often taken\footnote{Some references indeed state that synchronization to an inertial reference frame, like the \gls{bcrs}, is \textit{required} to perform \gls{tdi} (e.g., \cite{Tinto:2003vj,pireaux:2007sh}), while others suggest to synchronize all data to a virtual \textit{constellation time}, for example given by one of the spacecraft clocks (e.g., \cite{Wang:2014zba,Tinto:2020fcc}).} as the \gls{tcb}, such that all quantities appearing in the standard \gls{tdi} formulation are implicitly expressed as functions of the \gls{tcb}.

The total phase\footnote{The formulation of \gls{tdi} in total phase is not standard in the literature, which typically assumes large phase trends are removed from the data before processing (e.g., \cite{Tinto:2020fcc,hellings:2001aa,PhysRevD.98.042003,Otto:2012dk}).
} $\Phi_i(t)$ of the light originating from a laser source is given in the \gls{tcb} as
\begin{equation}
    \Phi_i(t) = \omega_i t + \phi_i(t)
    \qc
\label{eq:total-phase-tcb}
\end{equation}
where $\omega_i$ is the  laser frequency and $\phi_i(t)$ is the laser phase fluctuation, both expressed in \gls{bcrs} coordinates.

The phase difference $\eta_{ij}(t)$, for $i,j = 1,2,3$, measures the interference between the laser light received from spacecraft~$j$ and the local laser on spacecraft~$i$, as a function of the \gls{tcb}. Formally,
\begin{equation}
    \eta_{ij}(t) = \Phi_j(t - d_{ij}(t)) - \Phi_i(t)
    \qs
\label{eq:eta-tcb}
\end{equation}
Here, we define the 6 \glspl{ltt} $d_{ij}(t)$ such that 
\begin{equation}
     t - d_{ij}(t)
\end{equation}
is the emission time of the light emitted by spacecraft~$j$, which is received on spacecraft~$i$ at time $t$, where both times are expressed in the \gls{tcb}.

We introduce the delay operator $\delay{ij}$ defined in terms of \glspl{ltt}, which acts on any function $f(t)$ as
\begin{equation}
    \delay{ij} f(t) = f(t - d_{ij}(t))
    \qs
\label{eq:ltt-delay-operator}
\end{equation}
\Cref{eq:eta-tcb} now reads
\begin{equation}
    \eta_{ij}(t) = \delay{ij} \Phi_j(t) - \Phi_i(t)
    \qs
\label{eq:eta-with-delay-operators}
\end{equation}
We define the usual shorthand notation for chained delay operators\footnote{For two delays, we get $\delay{ijk} = \delay{ij} \delay{jk}$, which can be expanded to $\delay{ijk} f(t) = f(t - d_{ij}(t) - d_{jk}(t - d_{ij}(t)))$.}, 
\begin{equation}
    \delay{i_1 i_2 \dots i_n} = \delay{i_1 i_2} \delay{i_2 i_3} \dots \delay{i_{n-1} i_n}
    \qc
\end{equation}
which we use to write the first-generation \gls{tdi} Michelson combination $X_1$ \cite{Armstrong:1999hp} as
\begin{equation}
\begin{split}
    X_1(t) &= (1 - \delay{121})(\eta_{13}(t) + \delay{13} \eta_{31}(t)) \\
    &\qquad - (1 - \delay{131}) (\eta_{12}(t) + \delay{12} \eta_{21}(t))
    \qs
\end{split}
\label{eq:standard-X1}
\end{equation}
Here, all measurements are expressed in terms of the \gls{tcb} and all delay operators are defined in terms of \glspl{ltt}. The two other Michelson combinations $Y_1$ and $Z_1$ can be deduced by circular permutation of the indices.

Inserting \cref{eq:eta-with-delay-operators} in \cref{eq:standard-X1} yields the residual in $X_1$ that can be expressed in terms of the commutator of delay operators~\cite{Bayle:2018hnm},
\begin{equation}
    X_1(t) = \delay{13121} \Phi_1(t) - \delay{12131} \Phi_1(t)
    \qs
\label{eq:ltt-X1-residuals}
\end{equation}
In the case of a static constellation, the delay operators commute, such that laser noise cancels exactly in $X_1$.

Accounting for the flexing of the constellation, the overall delays applied by $\delay{13121}$ and $\delay{12131}$ differ by a small time interval $\Delta t_{X_1}$, which physically corresponds to the mismatch in the emission times of the two interfering virtual laser beams (here expressed in the \gls{tcb}). We can approximate its value by assuming our travel times to be linear functions of the form $d_{ij}(t) = \bar d + \delta d_{ij} + \dot d_{ij} t$, with $\bar d\approx \SI{8.3}{\second}$ as the average armlength and $\delta d_{ij}$, $\dot d_{ij}$ as small constants, and expanding 
\cref{eq:ltt-X1-residuals} to first order in $\delta d_{ij}$ and $\dot d_{ij}$. This gives\footnote{This expression is sometimes found in the literature simplified to $4 \bar d(\dot d_{13} - \dot d_{12})$ (e.g., \cite{hartwig-thesis}), assuming $\dot d_{ij} = \dot d_{ji}$. While this approximation is valid for the \gls{ltt} in the \gls{tcb}, it is not valid for the pseudoranges considered in the next section, in which potentially large clock errors enter with opposite sign in $\dot d_{ij}$ and $\dot d_{ji}$.}
\begin{equation}
    \Delta t_{X_1} \approx 2 \bar d(\dot d_{13} + \dot d_{31} - \dot d_{12} - \dot d_{21})\qs
\end{equation}
This evaluates to $\Delta t_{X_1} \approx \SI{E-7}{\second}$ (see also, e.g.,~\cite{Muratore:2020mdf}), and varies on annual timescales that are related to the orbital motion of the constellation around the Sun. We can assume that $\Delta t_{X_1}$ remains constant for the purpose of estimating the residual noise in a \gls{tdi} combination that involves time intervals of order $4 \bar d \approx \SI{33}{\second}$.

We expand \cref{eq:ltt-X1-residuals} to first order in $\Delta t_{X_1}$,
\begin{subequations}
\begin{align}
    X_1(t) &\approx \delay{12131} \Delta t_{X_1} \dot \Phi_1(t)
    \\
    &\approx \Delta t_{X_1} \omega_1 + \delay{12131} \Delta t_{X_1} \dot \phi_1(t)
    \qs
\end{align}
\end{subequations}
The first term $\Delta t_{X_1} \omega_1$ is outside the \gls{lisa} measurement band ($> \SI{E-5}{\hertz}$) and we neglect it. The second term is the residual laser frequency noise in $X_1$, which is above requirements~\cite{Tinto:2003vj}.

Therefore, one has to rely on the second-generation Michelson combinations $X_2$ for all practical purposes \cite{tdi-2-shaddock,Tinto:2003vj}, which has been shown to suppress laser noise to the required levels~\cite{Bayle:2018hnm},
\begin{equation}
\begin{split}
    X_2(t) &= (1 - \delay{13121}) [\eta_{12}(t) + \delay{12} \eta_{21}(t) \\
    &\qquad\qquad+ \delay{121} (\eta_{13}(t) + \delay{13} \eta_{31}(t))]
    \\
    &- (1 - \delay{12131}) [\eta_{13}(t) + \delay{13} \eta_{31}(t)
    \\
    &\qquad\qquad+ \delay{131} (\eta_{12}(t) + \delay{12} \eta_{21}(t))]
    \qs
\end{split}
\label{eq:tdi-X2}
\end{equation}
For reference, the residual in $X_2$ reads
\begin{equation}
    X_2(t) = \delay{121313121} \Phi_1(t) - \delay{131212131} \Phi_1(t)
    \qs
\end{equation}
The residual corresponds again to a small delay difference $\Delta t_{X_2}$. To estimate it, we now have to also consider higher order terms, and assume our \gls{ltt} to be given as $d_{ij}(t) = \bar d + \delta d_{ij} + \dot d_{ij} t + \ddot d_{ij}t^2/2$. Expanding to leading orders now gives
\begin{equation}
\begin{split}
    \Delta t_{X_2} &\approx 2 \bar d \Big[ \qty(\dot d_{12} + \dot d_{21})^2 - \qty(\dot d_{13} + \dot d_{31})^2 \\
    & \qquad + 4 \bar d(\ddot d_{12} + \ddot d_{21} - \ddot d_{13} - \ddot d_{31})\Big]\qc
\end{split}
\end{equation}
which evaluates to $\Delta t_{X_2} \approx \SI{E-12}{\second}$, as also verified by the numerical calculations in \cite{Muratore:2020mdf}. We again expand the residual in $X_2$ in this small time interval, which yields improved laser noise suppression compared to $X_1$, sufficient to fulfill the \gls{lisa} requirements.

\subsection{TDI with pseudoranges and raw measurements}%
\label{sub:procrastinating-tdi}

In a more realistic picture of the instrument, each spacecraft is equipped with its own clock, which we label after the hosting spacecraft. Because the clocks are not synchronized to each other or to a common time frame, they will drift and jitter due to instrumental imperfections and relativistic effects. In the following, we call the time difference between the  clocks and \gls{tcb} \textit{clock deviations}, see \cref{eq:clock_dev}.

Because one cannot access the \gls{tcb} time scale on board the spacecraft, one cannot directly measure the \glspl{ltt}. Instead, phase modulations of the lasers allow on-board measurements of the pseudoranges $d^{\hat\tau_i}_{ij}(\tau)$. These are defined as the difference between the times of emission and reception of the laser light as given by the receiver and emitter spacecraft clocks, respectively. The pseudoranges are functions of the reception time $\tau$ on the receiving spacecraft, such that
\begin{equation}
    \tau - d^{\hat\tau_i}_{ij}(\tau)
\end{equation}
is the time shown on clock $j$ when the beam was emitted from spacecraft~$j$. Expressed in a global frame, the pseudorange is a combination of the associated \gls{ltt} $d_{ij}(t)$ and the clock deviations~$i$ and~$j$. In practice, the pseudoranges are measured by the optical metrology system of \gls{lisa}, which introduces additional ranging noise. In this section, we neglect this noise and discuss its effect in \cref{sec:prn}.

Because the pseudorange $d^{\hat\tau_i}_{ij}(\tau)$ is measured on spacecraft $i$, it is expressed as a function of the receiver clock~$i$, as indicated by the superscript $\hat\tau_i$.

Similarly, phase differences are measured according to the onboard clocks, such that they are given as functions of three different timescales $\hat\tau_i$, for $i=1,2,3$. We denote them $\eta^{\hat\tau_i}_{ij}(\tau)$.

\begin{figure}
    \centering
    \includegraphics[width=\columnwidth]{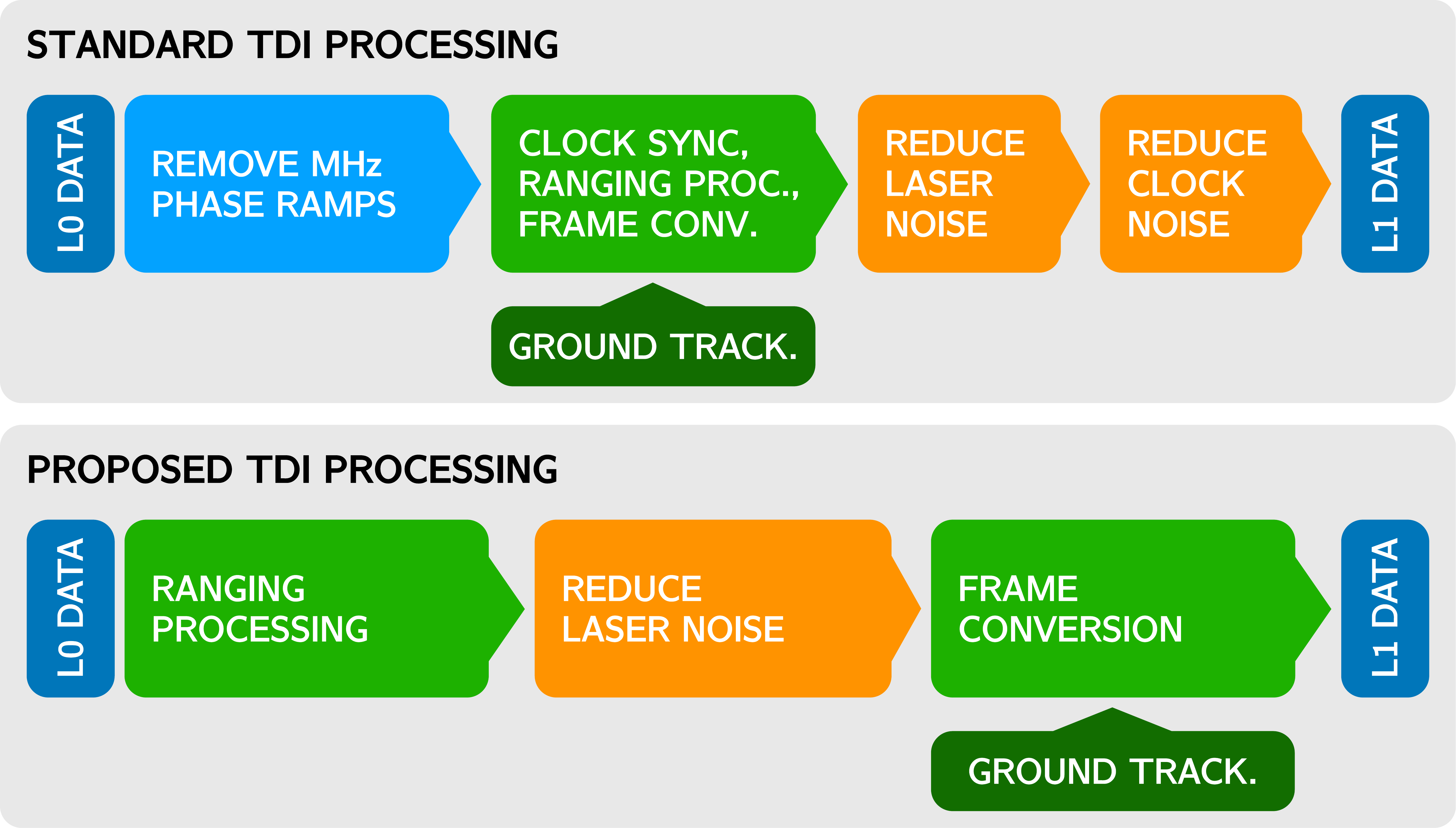}
    \caption{Comparison of standard noise reduction pipeline known from the literature and the alternative presented in this paper. We omit other noise suppression steps, such as the removal of spacecraft jitters.}
    \label{fig:pipelines}
\end{figure}

The approach documented in the literature to account for these effects is depicted in \cref{fig:pipelines}. It uses a Kalman-like filter to separate the \glspl{ltt} from the clock errors, given the pseudoranges, the ground-based estimates of the spacecraft positions, and the clock desynchronizations~\cite{Wang:2014zba,Wang:2015kja}. The clock error estimates are then used to synchronize all measurements to the \gls{tcb}. Finally, the re-synchronized measurements are used with the deduced \glspl{ltt} to perform the standard \gls{tdi} described in \cref{sub:standard-tdi}. This approach also includes the removal of \si{\mega\Hz} phase ramps and an additional step to correct for the residual clock noise~\cite{Hartwig:2020tdu}.

We show in this section that \gls{tdi} actually reduces laser noise to acceptable levels without the need to synchronize measurements to a common timescale (i.e., directly using the \textit{raw} measurements) and without the need to separate \glspl{ltt} and clock errors (i.e., directly using the measured pseudoranges). Furthermore, we show that this approach requires neither the removal of large trends from the beatnotes, nor the associated additional clock correction step at the end of the pipeline. This is summarized in the bottom panel of \cref{fig:pipelines}.

Let us re-express the interference between light received from spacecraft~$j$ and the local laser on spacecraft~$i$, introduced in \cref{eq:eta-tcb}, according to clock~$i$,
\begin{equation}
    \eta^{\hat\tau_i}_{ij}(\tau) = \Phi^{\hat\tau_j}_j(\tau - d^{\hat\tau_i}_{ij}(\tau)) - \Phi^{\hat\tau_i}_i(\tau)
    \qc
\label{eq:eta-clock-time}
\end{equation}
where the laser source phases $\Phi^{\hat\tau_i}_i(\tau)$ and $\Phi^{\hat\tau_j}_j(\tau)$ are now functions of the clock times~$i$ and~$j$, respectively.  Note that the value of $\tau$ in the previous equation is given in the clock time frame $\hat\tau_i$, and thus, when expressed in a global frame, includes the same clock drift and jitter that enter to the pseudorange $d^{\hat\tau_i}_{ij}(\tau)$.

Let us define a new delay operator $\delay{ij}^{\hat\tau_i}$ in terms of the pseudorange $d^{\hat\tau_i}_{ij}$ such that, applied to a quantity $f^{\hat\tau_j}(\tau)$ taken at a time measured on clock~$i$, it yields the same quantity evaluated at the time of emission by spacecraft~$j$. Formally,
\begin{equation}
    \delay{ij}^{\hat\tau_i} f^{\hat\tau_j}(\tau) = f^{\hat\tau_j}(\tau - d^{\hat\tau_i}_{ij}(\tau))
    \qc
\label{eq:pseudorange-operator}
\end{equation}
for any function $f^{\hat\tau_j}(\tau)$.

Note that the pseudorange delay operator $\delay{ij}^{\hat\tau_i}$ takes as input a function of the clock time indicated by the right index~$j$ and turns it into a function of the clock time corresponding to the left index~$i$. As a consequence, the pseudorange delay operator now performs two actions: it computes the time of emission of light as a function of the time of reception, expressed in different time frames, and also includes the time frame transformation between clock times $\hat\tau_j$ and $\hat\tau_i$ we use to describe the laser phases at these events.

These pseudorange delay operators can be chained using the usual shorthand notation; e.g., with two delays,
\begin{equation}
    \delay{ijk}^{\hat\tau_i} = \delay{ij}^{\hat\tau_i} \delay{jk}^{\hat\tau_j}
    \qc
\end{equation}
which can be expanded to
\begin{equation}
    \delay{ijk}^{\hat\tau_i} f^{\hat\tau_k}(\tau) = f^{\hat\tau_k}(\tau - d_{ij}^{\hat\tau_i}(\tau) - d_{jk}^{\hat\tau_j}(\tau - d_{ij}^{\hat\tau_i}(\tau)))
    \qs
\label{eq:chained-delays}
\end{equation}
The frame transformation rule still applies, as $ \delay{ijk}^{\hat\tau_i} f^{\hat\tau_k}$ is expressed in terms of clock time~$i$ while $f^{\hat\tau_k}$ is a function of clock time~$k$. More generally, chained delay operators convert clock time frames from their right-most index to their left-most index.

With this in mind, rewriting \cref{eq:eta-clock-time} using pseudorange delay operators makes it clearer that all quantities are measured in the frame of clock~$i$,
\begin{equation}
    \eta^{\hat\tau_i}_{ij}(\tau) = \delay{ij}^{\hat\tau_i} \Phi^{\hat\tau_j}_j(\tau) - \Phi^{\hat\tau_i}_i(\tau)
    \qs
\label{eq:eta-pseudorange-operators}
\end{equation}

Similarly, Michelson \gls{tdi} combinations can be rewritten using pseudorange delay operators and measurements taken according to their respective clocks,
\begin{equation}
\begin{split}
    X_1^{\hat\tau_1}(\tau) &= (1 - \delay{121}^{\hat\tau_1})(\eta_{13}^{\hat\tau_1}(\tau) + \delay{13}^{\hat\tau_1} \eta_{31}^{\hat\tau_3}(\tau)) \\
    &\qquad - (1 - \delay{131}^{\hat\tau_1}) (\eta_{12}^{\hat\tau_1}(\tau) + \delay{12}^{\hat\tau_1} \eta_{21}^{\hat\tau_2}(\tau))
    \qs
\end{split}
\label{eq:X1-clock-time}
\end{equation}
Note that all quantities appearing in $X_1$ are functions of clock time~1, such that the entire combination is a function of the same time frame.  The \gls{tdi} variable $X_1$ therefore bears the superscript $\hat\tau_1$.

\Cref{eq:eta-pseudorange-operators,eq:X1-clock-time} have the same algebraic structure as the standard \gls{tdi} \cref{eq:eta-with-delay-operators,eq:standard-X1}. Therefore, the same algebraic results hold, and we get a residual term in $X_1^{\hat\tau_1}$ that is the commutator of pseudorange delay operators similar to \cref{eq:ltt-X1-residuals},
\begin{equation}
    X_1^{\hat\tau_1}(\tau) = \delay{13121}^{\hat\tau_1} \Phi_1^{\hat\tau_1}(\tau) - \delay{12131}^{\hat\tau_1} \Phi_1^{\hat\tau_1}(\tau)
    \qs
    \label{eq:X1-in-clock-time}
\end{equation}
The physical interpretation of this equation is identical to that of \cref{eq:ltt-X1-residuals}, as the time interval $\Delta t^{\hat\tau_1}_{X_1}$ between the delays applied by $\delay{13121}^{\hat\tau_1}$ and $\delay{12131}^{\hat\tau_1}$ still corresponds to the mismatch in the emission times of the two virtual beams that make up the interferometric measurement, now expressed according to clock~1. The residual $X_1$ corresponds to the accumulated phase noise of the laser on spacecraft~1 between the two emission events. As such, it is invariant with respect to the time frame used to express it. In other words, 
\begin{equation}
    X_1^{\hat\tau_1}(\tau) = X_1^t (t^{\hat\tau_1}(\tau))
    \qs
    \label{eq:eq_X1}
\end{equation}
The full demonstration\footnote{The previous equation is exact when one considers phase measurements. In the case of frequency measurements, there is formally an additional scale factor of the order $(1 + \num{e-7})$ due to the derivative of the clock deviations, which does not significantly impact the residual noise level.} can be found in \cref{sec:eq_X1}. We conclude that laser noise is suppressed to the same level as in \cref{sub:standard-tdi}.

The same argument applies identically to the second-generation $X_2^{\hat\tau_1}$, where we immediately recover
\begin{equation}
    X_2^{\hat\tau_1}(\tau) = \delay{121313121}^{\hat\tau_1} \Phi_1^{\hat\tau_1}(\tau) -  \delay{131212131}^{\hat\tau_1} \Phi_1^{\hat\tau_1}(\tau) 
    \qs
\end{equation}
We conclude that $X_2^{\hat\tau_1}$ reduces laser noise to required levels, as well as the circular permutations $Y_2^{\hat\tau_2}$ and $Z_2^{\hat\tau_3}$. This is also demonstrated numerically in \cref{sec:sim_results}.

Note, however, that  $X_2^{\hat\tau_1}$, $Y_2^{\hat\tau_2}$, and $Z_2^{\hat\tau_3}$ are still defined according to different clocks. They may need to be synchronized to a global time frame, such as \gls{tcb} before they can be used for further scientific analysis. Since laser noise is already suppressed, the required timing precision for this final synchronization is now driven only by the needs of the astrophysical data analysis, much less stringent [M. Vallisneri and A. Petiteau, personal communication, 2021].

\section{Simulation model and TDI variable construction}%
\label{sec:sim_model}

\subsection{Interferometric measurements}%
\label{ssec:ifo_measurements}

To validate the alternative algorithm proposed in this paper and check that one can reach the required noise levels using the raw unsynchronized data, we rely on numerical simulations. These simulations are performed with the latest versions of \texttt{LISA Instrument}~\cite{lisainstrument} and \texttt{PyTDI}~\cite{pytdi}.  In contrast with the theoretical description of \cref{sec:theory}, which describes the model in terms of phase, the actual simulations are performed in frequency.

A detailed description of the model implemented in \texttt{LISA Instrument} is given in~\cite{Bayle:2019dfu,hartwig-thesis}.  Note that \texttt{LISA Instrument} uses two separate quantities to describe large out-of-band drifts and small in-band fluctuations. Here, we provide a simplified model that is directly formulated in terms of total frequency and only includes the instrumental effects relevant for this study.

\begin{figure}
    \centering
    \includegraphics[width=\columnwidth]{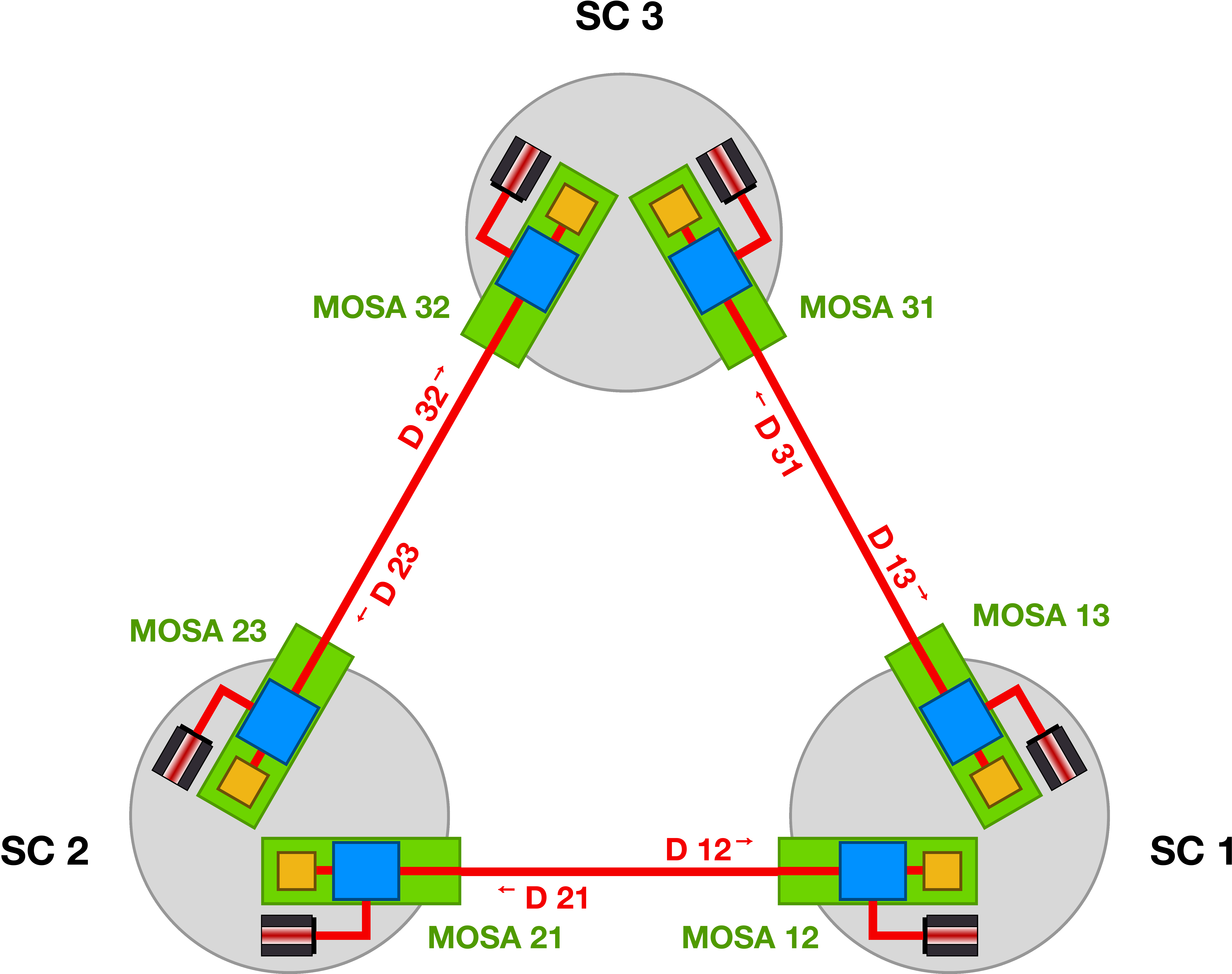}
    \caption{Model of the instrument as seen when looking down at the solar panels, with a total of 6 laser sources.}
    \label{fig:indexing6}
\end{figure}

On each spacecraft, we consider 2 laser sources\footnote{So far, we have considered only 1 laser per spacecraft for simplicity, but without loss of generality. The simulations are performed using a more realistic instrumental setup with 2 lasers per spacecraft.}, see \cref{fig:indexing6}, producing a total of 6 laser beams. For each beam, we model separately the total carrier frequency $\nu_{ij}^{\hat\tau_i}(\tau)$ and the upper sideband frequency $\nu_{ij,sb}^{\hat\tau_i}(\tau)$, where the indices $ij$ refer to the \gls{mosa} from which the beam is emitted. The carrier frequency of each beam is within a few \si{\mega\hertz} of the central frequency, defined as \SI{281.6}{\tera\hertz}. The sidebands are generated by modulating the carrier with a clock-derived signal at either $\nu_m = \SI{2.4}{\giga\hertz}$ or $\nu_m = \SI{2.401}{\giga\hertz}$ for the left- and right-hand side optical benches, respectively. In the end, we have
\begin{align}
    \nu_{ij,sb}^{\hat\tau_i}(\tau) = \nu_{ij}^{\hat\tau_i}(\tau) + \nu_m(1 + M_{ij}^{\hat\tau_i}(\tau))
    \qs
\end{align}

Here, $M_{ij}^{\hat\tau_i}(\tau)$ is a noise term accounting for imperfections in the modulation\footnote{\label{foot:modulation-levels}The optical sidebands are generated from the \si{\mega\hertz} clock tones by electrically up-converting them to \SI{2.4}{\giga\hertz} and \SI{2.401}{\giga\hertz} before modulating them onto the beams. Following~\cite{Barke:2015srr}, we expect the \SI{2.4}{\giga\hertz} sidebands to be dominated by noise due to fibre amplifiers, while the \SI{2.401}{\giga\hertz} sidebands are expected to be limited by the noisier electrical frequency conversion chain.}. Since we model these signals in their respective clock frames, $M_{ij}^{\hat\tau_i}(\tau)$ does \textit{not} contain any imperfections of the clocks themselves.

Each spacecraft carries 2 optical benches, on each of which beams are combined in 3 different interferometers. One of these interferometers is responsible for the test-mass readout, and is not relevant in this study. We include it in the following for completeness\footnote{However, for clarity, we do not include the frequency shifts due to the spacecraft motion, which are suppressed using the test-mass interferometer.}.

The \textbf{inter-spacecraft interferometers} yields the following carrier and sideband beatnotes
\begin{subequations}\label{eq:isc-beatnote-all}
\begin{align}
    \isc_{ij}^{\hat\tau_i}(\tau) &= \ddelay{ij}^{\hat\tau_i}\nu_{ji}^{\hat\tau_j}(\tau) - \nu_{ij}^{\hat\tau_i}(\tau) + N_{ij}^\isc(\tau)
    \qc
\label{eq:isc-beatnote}
    \\
    \isc_{ij,sb}^{\hat\tau_i}(\tau) &= \ddelay{ij}^{\hat\tau_i}\nu_{ji,sb}^{\hat\tau_j}(\tau) - \nu_{ij,sb}^{\hat\tau_i}(\tau)+ N_{ij}^{\isc,sb}(\tau)
    \qs
\label{eq:isc-beatnote2}
\end{align}
\end{subequations}
The \textbf{reference interferometers} beatnotes read
\begin{subequations}\label{eq:ref-beatnote-all}
\begin{align}
    \rfi_{ij}^{\hat\tau_i}(\tau) &= \nu_{ik}^{\hat\tau_i}(\tau) - \nu_{ij}^{\hat\tau_i}(\tau)+ N_{ij}^\rfi(\tau)
    \qc
\label{eq:ref-beatnote}
    \\
    \rfi_{ij,sb}^{\hat\tau_i}(\tau) &= \nu_{ik,sb}^{\hat\tau_i}(\tau) - \nu_{ij,sb}^{\hat\tau_i}(\tau) + N_{ij}^{\rfi,sb}(\tau)
    \qc
\end{align}
\end{subequations}
while those of the \textbf{test-mass interferometer} beatnotes are given as
\begin{equation}
    \tmi_{ij}^{\hat\tau_i}(\tau) = \nu_{ik}^{\hat\tau_i}(\tau) - \nu_{ij}^{\hat\tau_i}(\tau) + N_{ij}^\tmi(\tau) - 2 N_{ij}^\delta(\tau)
    \qs
\label{eq:tmi-beatnote}
\end{equation}
Here, the indices $i,j,k$ are to be chosen from the set $\{1,2,3\}$, with $i\neq j \neq k$. The $N_{ij}^\isc(\tau)$, $N_{ij}^\rfi(\tau)$ and $N_{ij}^\tmi(\tau)$ terms denote an uncorrelated readout noise in each interferometer, while $N_{ij}^\delta(\tau)$ is the frequency shift due to test-mass displacement. We define the Doppler-delay~\cite{Bayle:2021mue} $\ddelay{ij}^{\hat\tau_i}$ as
\begin{equation}
    \ddelay{ij}^{\hat\tau_i} f(\tau) = (1 - \dot d_{ij}^{\hat\tau_i}(\tau)) f(\tau - d_{ij}^{\hat\tau_i}(\tau))
    \qs
\label{eq:doppler-delay}
\end{equation}

In addition to the \si{\giga\hertz} sidebands, each laser beam is modulated with a \gls{prn} code, which allows an absolute measurement of the pseudoranges. We model this as a direct measurement of the clock time difference
\begin{equation}
    \prn_{ij}^{\hat\tau_i}(\tau) = d_{ij}^{\hat\tau_i}(\tau) + N_{ij}^\prn(\tau) + B_{ij}
    \qs
\label{eq:prn-with-bias}
\end{equation}
Note that this \gls{prn} measurement carries a relatively large ranging noise $N_{ij}^\prn$, which we assume to have zero mean. We include the term $B_{ij}$ to account for a potential constant bias in each pseudorange measurement.

\subsection{TDI combinations}
\label{ssec:tdi-vars}

We perform \gls{tdi} processing using frequency data following the prescriptions of~\cite{Bayle:2021mue}, using only Doppler delay operators.

However, neither the pseudorange $d_{ij}^{\hat\tau_i}(\tau)$ nor its derivative $\dot d_{ij}^{\hat\tau_i}(\tau)$ are known exactly. Therefore, they must be estimated from the onboard measurements. We apply these Doppler delays on total frequency data measured in the different interferometers, which are of large magnitude (around \SI{10}{\mega\hertz}), but carry only small frequency fluctuations (laser frequency noise around \SI{30}{\hertz\per\hertz\tothe{0.5}}). As such, the Doppler factor $\dot{d}_{ij}^{\hat\tau_i}(\tau)$ in \cref{eq:doppler-delay} needs to be known at a much higher precision than the actual delay $d_{ij}^{\hat\tau_i}(\tau)$. We discuss how to estimate both $\dot{d}_{ij}^{\hat\tau_i}(\tau)$ and $d_{ij}^{\hat\tau_i}(\tau)$ from the raw measurements in \cref{sec:ranging-processing}, and will denote these two separate estimates as $\tilde d_{ij}^{\hat\tau_i}(\tau)$ and $\dot{\bar{d}}_{ij}^{\hat\tau_i}(\tau)$, respectively. For now, we simply define the estimated Doppler delay
\begin{equation}
    \dtdidelay{ij}^{\hat\tau_i} f(\tau) = (1 - \dot{\bar{d}}_{ij}^{\hat\tau_i}(\tau)) f(\tau - \tilde d_{ij}^{\hat\tau_i}(\tau))
    \qs
\label{eq:doppler-tdi-delay}
\end{equation}

Using these estimates, we follow the standard procedure to construct intermediary \gls{tdi} variables known from the literature. The first step removes spacecraft motion from the measurements (not included in our simulations, but described here for completeness),
\begin{equation}
\begin{split}
    \xi_{ij}^{\hat\tau_i}(t) &= \isc_{ij}^{\hat\tau_i}(t) +\frac{\rfi_{ij}^{\hat\tau_i}(t) - \tmi_{ij}^{\hat\tau_i}(t)}{2} \\
    &\qquad + \dtdidelay{ij}^{\hat\tau_i}\frac{\rfi_{ji}^{\hat\tau_j}(t) - \tmi_{ji}^{\hat\tau_j}(t)}{2} 
    \qs
\end{split}
\end{equation}
This correction only uses the difference between reference and test-mass interferometers onboard the same spacecrafts, which are nominally at the same frequency. Therefore, it does not significantly impact the analytical models for noise couplings described in \cref{sec:prn}.

Then, we remove the frequency fluctuations of the right-hand side lasers by constructing the intermediary variables $\eta_{ij}$, given by
\begin{subequations}
\begin{align}
    \eta_{12}^{\hat\tau_1}(t) &= \xi_{12}^{\hat\tau_1}(t) + \dtdidelay{12}^{\hat\tau_1}\frac{\rfi_{21}^{\hat\tau_2}(t) - \rfi_{23}^{\hat\tau_2}(t)}{2} 
    \qc
    \\
    \eta_{13}^{\hat\tau_1}(t) &= \xi_{13}^{\hat\tau_1}(t) + \frac{\rfi_{12}^{\hat\tau_1}(t) - \rfi_{13}^{\hat\tau_1}(t)}{2}
    \qc
\end{align}
\end{subequations}
while the remaining ones can be computed by cyclic permutations of the spacecraft indices.

Lastly, we construct the Michelson variable $X_2^{\hat\tau_1}$ using the following factorized form, particularly suited for numerical efficiency and stability,
\begin{equation}
\begin{split}
    \dot X_2^{\hat\tau_1}(t) &= (1 - \dtdidelay{13121}^{\hat\tau_1}) [\eta_{12}^{\hat\tau_1}(t) + \dtdidelay{12}^{\hat\tau_1} \eta_{21}^{\hat\tau_2}(t) \\
    &\qquad\qquad+ \dtdidelay{121}^{\hat\tau_1} (\eta_{13}^{\hat\tau_1}(t) + \dtdidelay{13}^{\hat\tau_1} \eta_{31}^{\hat\tau_3}(t))]
    \\
    &- (1 - \dtdidelay{12131}^{\hat\tau_1}) [\eta_{13}^{\hat\tau_1}(t) + \dtdidelay{13}^{\hat\tau_1} \eta_{31}^{\hat\tau_3}(t)
    \\
    &\qquad\qquad+ \dtdidelay{131}^{\hat\tau_1} (\eta_{12}^{\hat\tau_1}(t) + \dtdidelay{12}^{\hat\tau_1} \eta_{21}^{\hat\tau_2}(t))]
    \qs
\end{split}
\label{eq:X2-freq} 
\end{equation}

\section{Results and discussion}%
\label{sec:sim_results}

\begin{figure*}
    \centering
    \includegraphics[width=\textwidth]{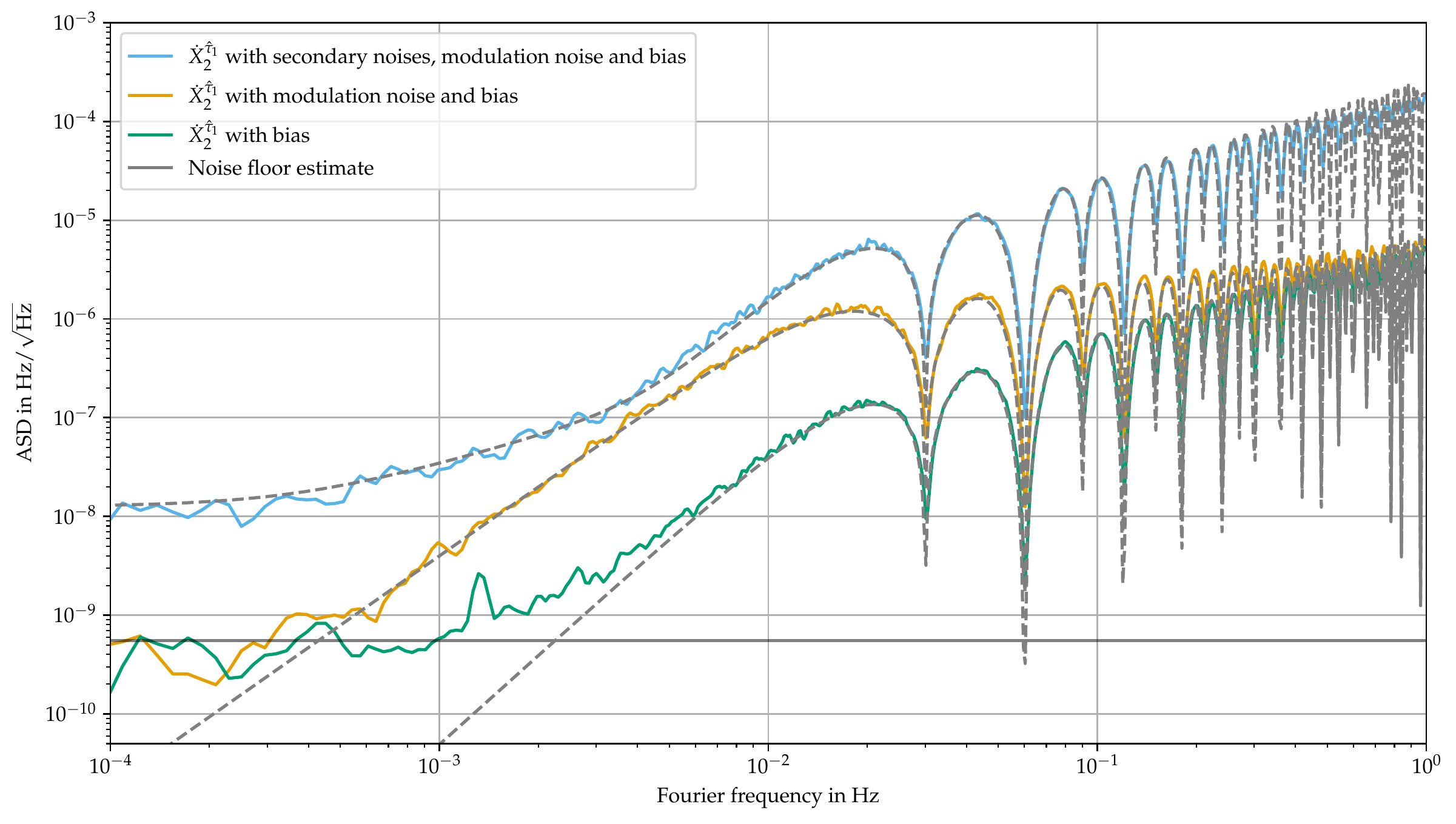}
    \caption{Residual noise levels in $\dot X_2^{\hat\tau_1}$ using the alternative version of the noise reduction pipeline presented in this paper. All simulations include laser frequency noise, \gls{prn} noise, as well as clock offsets, drifts, and jitters. The blue curve additionally includes test-mass acceleration, readout and modulation noises, as well as a deterministic bias in the \gls{prn}. Its level is well explained by the analytical model of the coupling of test-mass and readout noise into \gls{tdi}~\cite{hartwig-thesis}. The orange curve contains just the modulation noise and deterministic bias, while the green curve is generated with just the bias. Both are well explained by the analytical models presented in \cref{sec:prn} for most of the frequency band. At low frequencies, we hit a numerical noise floor, which we estimated by running another simulation with \textit{all} noises disabled (not shown). This noise-free simulation also contains the deviation from the analytical model and the small peak in the green curve slightly above \SI{e-3}{\hertz}. Therefore, we conclude that this deviation is a numerical artifact.
    \label{fig:sim-results1}}
\end{figure*}

In this section, we present the results of the alternative noise reduction pipeline described in \cref{ssec:tdi-vars}. Pseudoranges used in the pipeline are computed as given in \cref{sec:ranging-processing}, while the input data is simulated following \cref{ssec:ifo_measurements}.

All simulations have a duration of \SI{4e5}{\second}, and use orbits provided by \gls{esa}~\cite{Martens:2021phh,lisaorbits}. We perform three simulations that include different noise sources, and show that the results are in good agreement with the analytical models derived in \cref{sec:prn} in all cases, across most of the frequency band. We perform a fourth simulation with all noises disabled to evaluate the numerical limits of our simulations.

\subsection{Noise levels}
\label{ssec:noise-levels}

The laser noise \gls{psd} is given by
\begin{equation} \label{eq:Snu}
    \psd{\nu}(f) = \qty(\SI{30}{\hertz\per\hertz\tothe{0.5}})^2
    \qs
\end{equation}

Instrumental clock errors between the onboard clock times $\hat\tau_i$ and the spacecraft proper times $\tau_i$ are modelled as
\begin{equation} \label{eq:ClockErrorsTotal}
\begin{split}
    \hat\tau_i^{\tau_i}(\tau) &= \tau + x_{i,0} + y_i \tau + \frac{\dot{y}_i}{2} \tau^2 \\
    &\qquad + \frac{\ddot{y}_i}{3} \tau^3 + \int_{\tau_0}^{\tau}{y_i^\epsilon(\tau')\dd{\tau'}}
    \qc
\end{split}
\end{equation}
where $y_i^\epsilon(\tau)$ is a fractional frequency fluctuation with \gls{psd} (we only keep the low frequency part of the more complete model detailed in \cref{sec:clock}),
\begin{equation} \label{eq:Sclock}
    \psd{y} = \qty(\SI{6.32E-14}{\per\hertz\tothe{0.5}})^2 \qty(\frac{f}{\si{\hertz}})^{-1}
    \qs
\end{equation}
$x_{i,0}$ is a randomly determined constant deterministic clock offsets around \SI{1}{\milli\second}, and $y_i$, $\dot{y}_i$, $\ddot{y}_i$ are constant deterministic clock drifts with (conservative) orders of magnitude
\begin{subequations} \label{eq:deterministic-clock-effects}
\begin{align}
    y_i &\approx \SI{5e-7}{} \qc
    \\
    \dot{y}_i &\approx \SI{1.4e-14}{\per\second} \qc
    \\
    \ddot{y}_i &\approx \SI{9e-23}{\per\second\squared}
    \qs
\end{align}
\end{subequations}

Ranging noise in the \gls{prn} measurements is given by
\begin{equation} \label{eq:Sprn}
    \psd{\prn} = (\SI{3e-9}{\second})^2 \si{\per\hertz}
    \qc
\end{equation}
and a randomly determined deterministic bias in each arm with order of magnitude
\begin{equation} \label{eq:bias-prn}
    B_{ij} \approx \SI{3E-9}{\second}
    \qs
\end{equation}

We assume that the modulation noise level has a \gls{psd} of 
\begin{equation}\label{eq:Smod_left}
    \psd{M}(f) = \qty(\SI{5.2E-14}{\per\hertz\tothe{0.5}})^2 \qty(\frac{f}{\si{\hertz}})^{2/3}
\end{equation}
for left-handed sidebands and
\begin{equation}\label{eq:Smod_right}
    \psd{M}(f) = \qty(\SI{5.2E-13}{\per\hertz\tothe{0.5}})^2 \qty(\frac{f}{\si{\hertz}})^{2/3}
\end{equation}
for right-handed sidebands (accounting for \cref{foot:modulation-levels}).

Lastly, we include non-suppressed secondary noises in the form of test-mass acceleration noise with a \gls{psd} of
\begin{equation} \label{eq:Stm}
    \psd{\text{TM}}(f) = (\SI{2.4}{\femto\meter\per\second\squared\per\hertz\tothe{0.5}})^2 \qty(1 + \qty(\frac{\SI{0.4}{\milli\hertz}}{f})^2)
    \qc
\end{equation}
and interferometric readout noises with the noise shape
\begin{equation} \label{eq:Sifo}
    \psd{\text{OMS}}(f) = \psd{\text{ifo}} \times \qty(1 + \qty(\frac{\SI{2}{\milli\hertz}}{f})^4)
    \qc
\end{equation}
where the noise levels $\psd{\text{ifo}}$ in the different interferometers are taken as 
\begin{subequations} \label{eq:Sifo_level}
\begin{align}
    \psd{\text{ifo}}^\isc &= (\SI{6.35}{\pico\meter\per\hertz\tothe{0.5}})^2
    \qc 
    \\
    \psd{\text{ifo}}^{\isc,sb} &= (\SI{12.5}{\pico\meter\per\hertz\tothe{0.5}})^2
    \qc 
    \\
    \psd{\text{ifo}}^\rfi &= (\SI{3.32}{\pico\meter\per\hertz\tothe{0.5}})^2
    \qc
    \\
    \psd{\text{ifo}}^{\rfi,sb} &= (\SI{7.90}{\pico\meter\per\hertz\tothe{0.5}})^2
    \qc
    \\
    \psd{\text{ifo}}^\tmi &= (\SI{1.42}{\pico\meter\per\hertz\tothe{0.5}})^2
    \qs
\end{align}
\end{subequations}

\subsection{Simulation results}

We show in \cref{fig:sim-results1} the \gls{asd} of our simulated $\dot X_2^{\hat\tau_1}$ data, estimated using the logarithmically-scaled \gls{psd} method~\cite{lpsd}. 

The solid blue line is the result of a simulation containing all effects and noise sources described in \cref{ssec:noise-levels}. The noise level is well explained by the analytical model (upper dashed grey line) of the two main non-suppressed noise source (test-mass and readout noises) coupling into \gls{tdi}, as described in the literature~\cite{hartwig-thesis}. This result demonstrates that all other noise sources included in our simulation are successfully suppressed below the required levels.

To reveal the effects that limit the clock and laser noise suppression, we show the results of two additional simulations. We first disable the test-mass and readout noises to obtain the orange line; this proves that modulation noise is the main limiting noise, and its level is well explained by the model given in \cref{sec:prn} and shown in the center dashed grey line.

For the green curve, we further disable modulation noise. This shows that the model for the residual laser noise coupling to a deterministic ranging bias (also given in \cref{sec:prn}, and shown in the lower dashed grey line) accurately predicts the remaining residual noise for frequencies above \SI{5}{\milli\hertz}. Below these frequencies, we are limited by numerical effects, which we estimate by running a simulation with \textit{all} noises disabled. This is illustrated by the solid black line, which is computed from the average \gls{asd} of the noise-free simulation between \SI{0.1}{\milli\hertz} and \SI{1}{\milli\hertz}. We do not show the full noise-free simulation results, for visual clarity of the plot, but remark that they also contain the structure visible in the green curve between \SI{1}{\milli\hertz} and \SI{5}{\milli\hertz}, which deviates from the analytical model. We conclude that this discrepancy from the expected value (lower dashed grey line) is explained by numerical artifacts. Note that these numerical effects are significantly below the limiting secondary noise sources that make up the blue curve.

To make sure that the model and results we present here are not affected by the $\si{\milli\second}$ clock offsets, we have run additional simulations with conservative clock offsets around \SI{1}{\second}. We find that the results are similar to those presented in \cref{fig:sim-results1}, and still match closely our model, which does not include clock offsets and drifts.

As a final remark, let us note that all three analytic models given by the grey dashed lines are not specific to the alternative pipeline proposed in this paper, but are equally valid for the \textit{standard} implementation found in the literature.

\section{Conclusion}%
\label{sec:conclusion}

The standard procedure to compute laser noise-suppressing \gls{tdi} combinations found in the literature requires an initial processing step to synchronize the beatnote measurements (initially recorded according to different clocks) to a common time frame, usually the \gls{tcb}, in which the scientific analysis can be carried out. In this paper, we highlighted the fact that the same laser noise reduction can be achieved without prior synchronization of the raw data streams, significantly simplifying the noise reduction pipeline and making it free of any potential artefacts introduced by clock synchronization and reference frame transformations. One must still transform the resulting laser-noise free combinations to express them in a time coordinate frame relevant for the scientific analysis. However, the required timing precision is much less stringent ($\sim \SI{1}{\milli\second}$) than the one required for laser noise reduction ($\sim \SI{10}{\nano\second}$) or clock-noise reduction ($\sim \SI{1}{\pico\second}$ \acrshort{rms}).

We formulated \gls{tdi} directly in the time frames of the onboard clocks, such that the usual light travel times that appear in the \gls{tdi} combinations are replaced by pseudoranges. These pseudoranges can be directly recovered from the onboard measurements, at a precision limited by the stability of the \si{\giga\hertz} sidebands. Since we only rely on local measurements, the required processing steps are much simpler than the previously suggested ranging processing~\cite{Wang:2014zba,Wang:2015kja}. Furthermore, this new approach operates directly on the \si{\mega\hertz} beatnotes given as the raw data, in contrast to the usual assumption that \gls{tdi} should be performed on some form of phase or frequency fluctuations around those beatnotes. This circumvents an additional clock-noise correction step, further simplifying the noise reduction pipeline. 

This approach of reducing laser and clock noise in the same processing step has also been recently been demonstrated in an experimental test bed~\cite{Yamamoto:2021ujg}, such that we are confident that it will be applicable to the real \gls{lisa} data.

We showed analytically and numerically that both laser and clock noises are successfully reduced to acceptable levels. The dominant limiting effect is due to errors in the sideband, often called modulation noise, which is also the case in the standard pipeline. We showed numerically that the impact of this limiting noise is indeed equivalent to the levels one can achieve with the standard algorithm, which uses phase or frequency fluctuations only, and requires an extra clock-noise reduction step.

While we only demonstrated this method for the widely used Michelson combinations, we expect it to be applicable to any \textit{geometric} combination, such as those found in~\cite{Vallisneri:2005ji,Muratore:2020mdf}. Combinations of these geometric combinations, such as the quasi-orthogonal $A$, $E$ and $T$ channels~\cite{Prince2002}, will require a prior synchronization of the base variables used to construct them, but again, with much looser timing requirements than that those required for laser-noise suppression.

As a continuation of the presented study, one could quantify the required accuracy of the final synchronization step in future work, both for the construction of variables like $A$, $E$ and $T$, but also for the scientific exploitation of the data. In addition, a direct quantitative comparison between the standard \gls{inrep} pipeline and our proposed alternative would be of interest for future studies. Other follow-up works could include other noise-reduction steps envisioned for the \gls{lisa} mission, such as the suppression of noise induced by longitudinal and angular spacecraft jitters.

\appendix

\section{Proof of equivalence of TDI processing in TCB and clock frame}%
\label{sec:eq_X1}

Let us demonstrate \cref{eq:eq_X1} analytically, i.e., that the formulation of \gls{tdi} using desynchronized measurements and pseudoranges is equivalent to formulating it in \gls{tcb}, with the exception that the final result is expressed in the local clock frame.

Let us recall that the \gls{ltt} $d_{ij}$ is the \gls{tcb} time taken by light to propagate from spacecraft~$j$ to spacecraft~$i$, see \cite{chauvineau:2005pr}. It is expressed as a function of the receiver time. In other words, $t-d_{ij}(t)$ is the \gls{tcb} at the emission of the signal, as a function of $t$, the \gls{tcb} at reception.

The pseudoranges $d^{\hat\tau_i}_{ij}$ correspond to the same quantities but expressed in terms of clock time $\hat\tau_i$. More precisely $\tau - d^{\hat\tau_i}_{ij}(\tau)$ is the clock time of spacecraft~$j$ at the instant of emission of the signal as a function of $\tau$,  the clock time of spacecraft~$i$ at the reception of the signal. We can formally express this emission time using the \gls{ltt} as
\begin{equation}
    \tau - d_{ij}^{\hat \tau_i}(\tau) = \hat\tau_j^t (t_i - d_{ij}(t_i))
    \qc 
\end{equation}
where we used the relationship between $\hat\tau_i$ and $t$ defined in \cref{sec:theory} and introduced the shorthand notation
 \begin{equation}
    t_i = t^{\hat\tau_i}(\tau)
    \qs
    \label{eq:ti}
\end{equation}
Let us now consider $F^{\hat\tau_i}(\tau)=\delay{ij}^{\hat\tau_i}f^{\hat\tau_j}(\tau)$, where $f$ is a function that characterizes an observable (i.e. an invariant quantity). Using the definition of the delay operator introduced in \cref{eq:pseudorange-operator} and \cref{eq:ti},
\begin{equation}
\begin{split}
    F^{\hat\tau_i}(\tau) &= f^{\hat\tau_j}\qty(\tau-d_{ij}^{\hat\tau_i}(\tau))
    \qc 
    \\
    &=f^{\hat\tau_j}\qty(\hat\tau_j^t\qty(t_i - d_{ij}(t_i)))
    \qc
    \\
    &=f^t\qty(t_i - d_{ij}(t_i))
    \qc
    \\
    &= \left[\delay{ij} f(t)\right]_{t=t^{\hat\tau_i}(\tau)}
    \qs
\end{split}
\end{equation}
This shows that 
\begin{equation}
    \delay{ij}^{\hat\tau_i}f^{\hat\tau_j}(\tau) = \delay{ij} f \qty(t^{\hat\tau_i}(\tau))
    \qs
\end{equation}
Applying this result iteratively leads to \cref{eq:eq_X1}.

\section{Ranging processing}
\label{sec:ranging-processing}

Let us describe how the Doppler factor $\dot{\bar{d}}_{ij}^{\hat\tau_i}(\tau)$ and the delay $\tilde d_{ij}^{\hat\tau_i}(\tau)$ in \cref{eq:doppler-tdi-delay} can be estimated.  

To obtain a first estimate of the Doppler factor, which we note $\dot{\bar{d}}_{ij}^{\hat\tau_i}(\tau)$, we combine the sideband and carrier beatnotes as follows,
\begin{subequations}
\begin{align}
    \dot{\bar{d}}_{12}^{\hat\tau_1}(\tau) &= \frac{\isc_{12}^{\hat\tau_1}(\tau) - \isc_{12, sb}^{\hat\tau_1}(\tau) + \SI{1}{\mega\hertz}}{\SI{2.401}{\giga\hertz}}
    \qc
    \\
    \dot{\bar{d}}_{13}^{\hat\tau_1}(\tau) &= \frac{\isc_{13}^{\hat\tau_1}(\tau) - \isc_{13, sb}^{\hat\tau_1}(\tau) - \SI{1}{\mega\hertz}}{\SI{2.4}{\giga\hertz}}
    \qc
\end{align}
\end{subequations}
for the left- and right-handed optical benches, respectively. Inserting the expressions of the inter-spacecraft beatnotes from \cref{eq:isc-beatnote,eq:isc-beatnote2} in those expressions yields
\begin{subequations}
\begin{align}
    \dot{\bar{d}}_{12}^{\hat\tau_1}(\tau) &= \dot d_{12}^{\hat\tau_1}(\tau) + \frac{\SI{2.4}{\giga\hertz}}{\SI{2.401}{\giga\hertz}} M_{12}^{\hat\tau_1}(\tau) - \ddelay{12}^{\hat\tau_1} M_{21}^{\hat\tau_2}(\tau)
    \qc
    \label{eq:tdi-doppler-factor-uncorrected1}
    \\
    \dot{\bar{d}}_{13}^{\hat\tau_1}(\tau) &= \dot d_{13}^{\hat\tau_1}(\tau) + \frac{\SI{2.401}{\giga\hertz}}{\SI{2.4}{\giga\hertz}} M_{13}^{\hat\tau_1}(\tau) - \ddelay{13}^{\hat\tau_1} M_{31}^{\hat\tau_3}(\tau)
    \qs
    \label{eq:tdi-doppler-factor-uncorrected2}
\end{align}
\label{eq:tdi-doppler-factor-uncorrected}
\end{subequations}
These combinations contain the desired delay derivative $\dot d_{ij}^{\hat\tau_i}(\tau)$, but have additional contributions due to modulation errors. As mentioned in \cref{foot:modulation-levels}, we expect the \SI{2.401}{\giga\hertz} sidebands to have a higher modulation noise level. We shall see below that we can further reduce the noise coming from these sidebands.

To start with, let us write down the delay $\tilde d_{ij}^{\hat\tau_i}(\tau)$ as the sum of a constant part $d_{ij,0}$ and a time-varying part $d_{ij,v}(\tau)$,
\begin{equation}\label{eq:delay-estimate}
    \tilde d_{ij}^{\hat\tau_i}(\tau) = d_{ij,0} + d_{ij,v}(\tau)
    \qs
\end{equation}

We can compute $d_{ij,v}(\tau)$ by integrating $\dot d_{ij}^{\hat\tau_i}(\tau)$ as given in \cref{eq:tdi-doppler-factor-uncorrected},
\begin{equation}
\begin{split}
    \tilde d_{ij,v}^{\hat\tau_i}&(\tau) = \int_{\tau_0}^\tau \dot{\bar{d}}_{ij}^{\hat\tau_1}(\tau') \dd \tau' 
    \\
    &\approx d_{ij}^{\hat\tau_i}(\tau) - d_{ij}^{\hat\tau_i}(\tau_0) 
    + \int_{\tau_0}^\tau \qty[M_{ij}^{\hat\tau_i}(\tau')
     - \ddelay{ij}^{\hat\tau_i} M_{ji}^{\hat\tau_j}(\tau')]\dd \tau'
    \qs
\end{split}
\label{eq:bar-d-v}
\end{equation}
The dominant noise in this expression comes from the integrated \SI{2.401}{\giga\hertz} sideband modulation noises\footnote{For clarity, we neglect the \si{\giga\hertz} frequency ratio scaling factor from \cref{eq:tdi-doppler-factor-uncorrected1}, since it does not significantly change the noise level.}, whose noise level can be computed from \cref{eq:Smod_right} to be
\begin{equation}
    \psd{\smallint{M}}(f) = \qty(\SI{8.3E-14}{\second\per\hertz\tothe{0.5}})^2 \qty(\frac{f}{\si{\hertz}})^{-4/3}
    \qc
\end{equation}
which is far below the noise level of the \gls{prn}, and sufficient for our purposes. Note that $\tilde d_{ij,v}^{\hat\tau_i}(\tau)$ is only known relative to the initial value $d_{ij}^{\hat\tau_i}(\tau_0)$.  We can estimate this offset by computing the time averaged
\begin{equation}\label{eq:PRN-const}
    \begin{split}
    \tilde d_{ij,0}^{\hat\tau_i} &= \text{avg}\qty[ \prn_{ij}^{\hat\tau_i}(\tau) - \tilde d_{ij,v}^{\hat\tau_i}(\tau) ] 
    \\
    &\approx d_{ij}(\tau_0) + B_{ij} + \text{avg}\qty[N_{ij}^\prn]
    \qc
    \end{split}
\end{equation}
where we have used \cref{eq:prn-with-bias,eq:bar-d-v} and have ignored the sub-dominant modulation noise terms in the second line, for clarity. The constant terms $d_{ij}(\tau_0)$ and $B_{ij}$ are unaffected by the averaging, and the latter remains as a systematic bias on our estimate. However, the noise term $N_{ij}^\prn$ is strongly suppressed, leaving an error that we estimate as
\begin{equation}
    \text{std}[N_{ij}^\prn]/\sqrt{N_s}
    \qc
\end{equation}
where $\text{std}[N_{ij}^\prn]$ is the standard deviation of the ranging noise and $N_s$ is the number of samples we average over. Assuming the same \gls{prn} noise as in \cref{sec:sim_results} ($\text{std}[N_{ij}^\prn] \approx \SI{3}{\nano\second}$) and knowing that a \SI{3}{\nano\second} bias is already below specifications (see \cref{fig:sim-results1}), averaging over only a few \gls{prn} measurements is sufficient.

Overall, the dominant contributions to the left- and right-hand sided estimates are given by
\begin{subequations}
\begin{align}
    \tilde d_{12}^{\hat\tau_1}(\tau) &\approx d_{12}^{\hat\tau_1}(\tau)+ B_{12}  
     - \int_{\tau_0}^\tau \ddelay{12} M_{21}^{\hat\tau_2}(\tau')
     \dd \tau' 
    \qc
    \\
    \tilde d_{13}^{\hat\tau_1}(\tau) &\approx d_{13}^{\hat\tau_1}(\tau) + B_{13}
    + \int_{\tau_0}^\tau  M_{13}^{\hat\tau_1}(\tau')
     \dd \tau' 
    \qs
\end{align}
\label{eq:delay-estimate-final}
\end{subequations}
While this is sufficient for the delay part of the Doppler-delay operator, we still need to suppress the \SI{2.401}{\giga\hertz} sidebands modulation noise in the signal combinations in \cref{eq:tdi-doppler-factor-uncorrected} used as multiplicative factors. We can remove them by adapting the signal combination proposed in~\cite{Hartwig:2020tdu} to include the additional \SI{1}{\mega\hertz} offset between the sidebands. Let us define
\begin{equation}
\begin{split}
    \Delta M_1 = \frac{\rfi_{13, sb}^{\hat\tau_1} - \rfi_{13}^{\hat\tau_1} + \SI{1}{\mega\hertz}}{2} - \frac{\rfi_{12, sb}^{\hat\tau_1} - \rfi_{12}^{\hat\tau_1} - \SI{1}{\mega\hertz}}{2}
\end{split}
    \label{eq:DeltaM}
\end{equation}
and cyclic for $\Delta M_2$ and $\Delta M_3$.  We can use these expressions to correct our previous estimate,
\begin{subequations}
\begin{align}
    \dot{\bar{d}}_{12,c}^{\hat\tau_1}(\tau) &= \dot{\bar{d}}_{12}^{\hat\tau_1}(\tau) - \frac{\dtdidelay{12}^{\hat\tau_1}\Delta M_2}{\SI{2.401}{\giga\hertz}}
    \qc
    \label{eq:d12c}
    \\
    \dot{\bar{d}}_{13,c}^{\hat\tau_1}(\tau) &= \dot{\bar{d}}_{13}^{\hat\tau_1}(\tau) + \frac{\Delta M_1}{\SI{2.4}{\giga\hertz}}
    \qc
    \label{eq:d13c}
\end{align}
\label{eq:dijc}
\end{subequations}
where the quantities $\dot{\bar{d}}_{ij}^{\hat\tau_1}(\tau)$ are given in \cref{eq:tdi-doppler-factor-uncorrected}. Note that the Doppler delays in \cref{eq:dijc} have to be computed using the uncorrected Doppler factor given in \cref{eq:tdi-doppler-factor-uncorrected1}, and the previously estimated delays from \cref{eq:delay-estimate}, both of which still contain the higher modulation noise. The procedure described here could also be applied iteratively to get a better estimate. However, we believe that a single iteration is sufficient, as $\Delta M_i$ is already a small enough correction such that second order terms are negligible.

Substituting the expressions of the reference beatnotes into \cref{eq:DeltaM}, and using the resulting expressions in \cref{eq:d12c,eq:d13c} we find the approximate expressions
\begin{subequations}
\begin{align}
    \dot{\bar{d}}_{12,c}^{\hat\tau_1}(\tau) &\approx \dot d_{12}^{\hat\tau_1}(\tau) + \frac{\SI{2.4}{\giga\hertz}}{\SI{2.401}{\giga\hertz}} ( M_{12}^{\hat\tau_1} - \ddelay{12}^{\hat\tau_1}M_{23}^{\hat\tau_2} )
    \qc
\label{eq:tdi-doppler-corrected1}
    \\
    \dot{\bar{d}}_{13,c}^{\hat\tau_1}(\tau) &\approx \dot{d}_{13}^{\hat\tau_1}(\tau) + M_{12}^{\hat\tau_1} - \ddelay{13}^{\hat\tau_1} M_{31}^{\hat\tau_3}
    \qc
\label{eq:tdi-doppler-corrected2}
\end{align}
\label{eq:tdi-doppler-corrected}
\end{subequations}
where we have assumed that $\dtdidelay{12}^{\hat\tau_1}\Delta M_2 \simeq \ddelay{12}^{\hat\tau_1}\Delta M_2$. From these expressions, we see that the modulation noise terms from the \SI{2.401}{\giga\Hz} sidebands have been removed.

\section{Pseudorange measurement uncertainty}%
\label{sec:prn}

We discuss here how errors in the pseudoranges couple into the final \gls{tdi} variable. We consider two effects. Firstly, the \SI{2.4}{\giga\hertz} sideband modulation noise terms $M^{\hat\tau_i}_{ij}$ appearing in \cref{eq:tdi-doppler-corrected} are the dominating noise terms in our final estimate of the Doppler factor. Secondly, the ranging bias $B_{ij}$ is the most significant contribution in our estimate of the delay as given in \cref{eq:delay-estimate-final}.

\subsection{Modulation noise coupling}
\label{sub:modulation-noise-coupling}

The Doppler factor applied by each estimated delay $\dtdidelay{ij}$ introduces modulation errors, as described in \cref{eq:tdi-doppler-corrected}. The dominant coupling of these error terms is due to the large \si{\mega\hertz} beatnote frequencies the Doppler shifts are applied to. To estimate this coupling, we can assume the beatnote frequencies to be constant,
\begin{equation}
    \isc_{ij}^{\hat\tau_1}(\tau) = a_{ij}
    \qcomma
    \rfi_{ij}^{\hat\tau_1}(\tau) = b_{ij}
    \qs
\end{equation}
Furthermore, to estimate the contribution of modulation noise from each estimated delay operator, we use \cref{eq:doppler-tdi-delay,eq:tdi-doppler-corrected} to define
\begin{subequations}
\begin{align}
    \dtdidelay{12}^Mf(\tau) = (1 - M_{12}^{\hat\tau_j} + \delay{}M_{23}^{\hat\tau_2}) \times \delay{} f(\tau)
    \qc
\label{eq:M_tdi-doppler-corrected1}
    \\
    \dtdidelay{13}^Mf(\tau) = (1 - M_{12}^{\hat\tau_3} + \delay{}M_{31}^{\hat\tau_2}) \times \delay{} f(\tau)
    \qc
\label{eq:M_tdi-doppler-corrected2}
\end{align}
\label{eq:M_tdi-doppler-corrected}
\end{subequations}
and cyclic, with $\delay{}f(\tau) = f(\tau - \bar d)$ applying an average delay of $\bar d\approx \SI{8.3}{\second}$. In these expressions, it is sufficient to approximate $\tdidelay{}$ by $\delay{}$, round $\SI{2.4}{\giga\hertz} / \SI{2.401}{\giga\hertz}$ to $1$ and neglect $\dot d_{ij}^{\hat\tau_i}$ corrections.

Using \cref{eq:tdi-doppler-corrected} and cyclic, we can compute the contributions of the modulation noise to the variables $\eta_{ij}$.  They read
\begin{align}
    \eta_{12}^M &= a_{12} - b_{23} (1 - M_{12}^{\hat\tau_1} + \delay{} M_{23}^{\hat\tau_2}) \qc
    \\
    \eta_{31}^M &= a_{31} - b_{12} (1 - M_{31}^{\hat\tau_3} + \delay{} M_{12}^{\hat\tau_1}) \qc
    \\
    \eta_{13}^M &= a_{13} + b_{12} \qc
    \\
    \eta_{21}^M &= a_{21} + b_{23} \qc
\end{align}
where we have used that $b_{ij} = -b_{ik}$.
Inserting these expressions into \cref{eq:X2-freq} and using \cref{eq:M_tdi-doppler-corrected}, we obtain the following expression,
\begin{equation}
\begin{split}
    \dot X_2^M &= (1 - \delay{}^2 - \delay{}^4 + \delay{}^6) [a_{21} \delay{} M_{23}^{\hat\tau_2} - a_{31} \delay{} M_{31}^{\hat\tau_3} \\
    &\qquad + (a_{12} - a_{13} + b_{12} (\delay{}^2 - 1))M_{12}^{\hat\tau_1}]
    \qc
\end{split}
\end{equation}
where second order terms in $M^{\hat\tau_i}_{ij}$ have been neglected. 
Assuming the three modulation noises are uncorrelated with identical \glspl{psd} $\psd{M}$ and zero mean, the \gls{psd} of $X_2$ due to modulation noise is given as
\begin{equation}
    \psd{\dot X_2}^M(\omega) = 16 \sin(2\omega \bar d)^2\sin(\omega \bar d)^2 A_{X_2}(\omega)\psd{M}(\omega)
    \qc
\end{equation}
with
\begin{equation}
\begin{split}
    &A_{X_2}(\omega) = (a_{12} - a_{13})^2 + a_{21}^2 + a_{31}^2
    \\
    &\qquad - 4 b_{12} (a_{12} - a_{13} - b_{12}) \sin[2](\omega \bar{d})
    \qs
\end{split}
\end{equation}

Note that this formula is identical to the modulation noise coupling described in~\cite{Hartwig:2020tdu}, where it was derived under the assumption that a clock correction is applied to the detrended variables, instead of directly using the total frequency. This result is expected, as the argument developed in~\cite{Hartwig:2020tdu} still holds: when expressed in the spacecraft proper time frames, both modulation and clock noise enter the \gls{mpr} identically. This means that any \gls{tdi} formulation that suppresses clock noise using the sidebands will simply replace it with the modulation noise. As a result, modulation noise has exactly the same impact in both scenarios, and any prior performance evaluations in ~\cite{Hartwig:2020tdu} are still valid.

\subsection{Ranging bias impact}

We can model the impact of a bias in the delays $\tdidelay{ij}$ applied as part of $\dtdidelay{ij}$ as
\begin{subequations}
\begin{align}
    \tdidelay{ij} f(\tau) &= f(\tau - d_{ij}^{\hat\tau_i}(\tau) - B_{ij}) \\
    &\approx f(\tau -  d_{ij}^{\hat\tau_i}(\tau)) -  B_{ij}\dot f(\tau -  d_{ij}^{\hat\tau_i}(\tau)) \\
    &= \delay{ij} f(\tau)  - B_{ij} \delay{ij} \dot f(\tau)
    \qs
\label{eq:ranging-bias}
\end{align}
\end{subequations}
A ranging bias couples to the main noise contributions in the interferometric beatnotes given in \cref{eq:isc-beatnote,eq:ref-beatnote}. This coupling is given by laser frequency noise regardless of whether synchronization is performed before \gls{tdi} processing or not, such that we expect to find the same results obtained in previous studies. Plugging \cref{eq:ranging-bias} in $X_2$ and assuming all six lasers to be uncorrelated and of equal \gls{psd} $S_{\nu}$, the residual noise due to ranging biases can be computed to be
\begin{equation}
\begin{split}
    S^{N^{\prn,\text{bias}}}_{\dot X_2}(\omega) &= 16 \sin^2(\omega \bar d)\sin^2(2\omega \bar d) \omega^2
    \\ 
    &\times (B_{12}^2 + B_{21}^2 + B_{13}^2 + B_{31}^2) S_{\nu}(\omega)
    \qs
\end{split}
\end{equation}

\section{Clock model}%
\label{sec:clock}

The satellite clocks are \glspl{uso} delivering time scales $\hat\tau_i$ that deviate from the local proper time $\tau_i$ by a time offset $x_{i,0}$, a stochastic frequency noise $y_i^\epsilon(\tau)$, and deterministic frequency drifts $\dot{y}_i$, $\dot{y}_i$, and $\ddot{y}_i$, at linear, quadratic and cubic order respectively, all of instrumental origin. Formally one has
\begin{equation}\label{equ:real_clock}
\begin{split}
    \hat\tau_i^{\tau_i}(\tau) &= \tau + x_{i,0} + y_i \tau + \frac{\dot{y}_i}{2} \tau^2 \\
    &\qquad + \frac{\ddot{y}_i}{3} \tau^3 + \int_{\tau_0}^{\tau}{y_i^\epsilon(\tau')\dd{\tau'}}
    \qc
\end{split}
\end{equation}

We assume that the three clocks are uncorrelated. Therefore, the deterministic and noise terms are independent, but of similar order of magnitude. If one is interested in the very long term (5 years or more) an additional quartic frequency drift term can be added to better fit the data. Here, we work under the assumption that the cubic model is sufficient.

The offset $x_{i,0}$ depends on the time at which the clock was switched on and can in principle take any value. Here, we assume that it is known and corrected to within about a millisecond using \gls{esa} ground tracking.

The linear term $y_i \tau$ is usually not specified. Assuming it is calibrated before launch and that about a year has passed between the time offset calibration and the switching on of the clock, one has $y_i \approx \num{5e-7}$, obtained by integrating the quadratic term (see below) for a year.

The coefficients of the quadratic and cubic terms are estimated using the \glspl{uso} on the Cassini and GRAIL missions (see, e.g., pages 9 and 23 of \cite{asmar:2012}), where the fractional frequency linear and quadratic drifts are of order \SI{4.8e-7}{\per yr} and \SI{-7.4e-8}{\per yr\squared} respectively, leading to the following orders of magnitudes for $\dot{y}_i$ and $\ddot{y}_i$,
\begin{subequations}
\begin{align}
    \abs{\dot{y}_i} &\approx \SI{1.6E-14}{\per\second}
    \qc
    \\
    \abs{\ddot{y}_i} &\approx \SI{9E-23}{\per\second\squared}
    \qs
\end{align}
\label{eq:C2}
\end{subequations}
The actual values and signs are specific to each clock and the clock's local environment (temperature, magnetic field, radiation, etc.) can also add additional dependencies to \cref{equ:real_clock}. The linear drift and \cref{eq:C2} can be corrected to some extent using \gls{esa} ground tracking, as for the time offset, but here we conservatively keep the typical uncorrected values in our model in order to reduce our reliance on any pre-processing of the raw data.

\begin{figure}[!]
    \centering
    \includegraphics[width=\columnwidth]{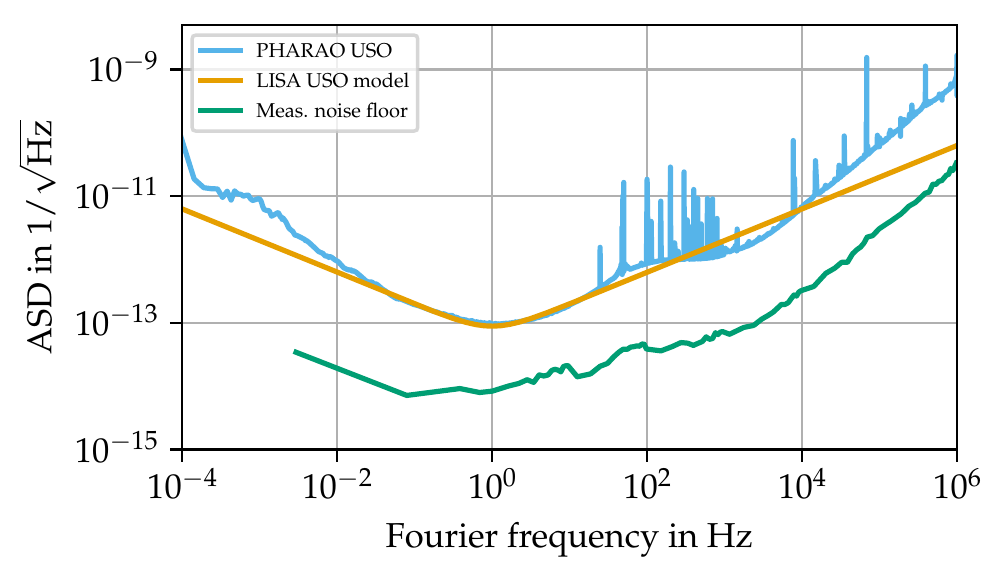}
    \caption{Comparison of the measured phase noise of the ACES/PHARAO \gls{uso} and the noise model from \cref{equ:USO_PSD}, expressed as fractional frequency fluctuations. The green curve is the estimated noise floor of the measurement device (Agilent 5125A). Measurement at \SI{100}{\mega\hertz} with the reference frequency provided by a cryogenic sapphire oscillator (ULISS), with negligible broadband noise but causing the spurious peaks visible in the graph.}
    \label{fig:PHARAO}
\end{figure}

The stochastic noise can be described by the sum of a flicker frequency noise and a flicker phase noise with an Allan variance (in fractional frequency) of roughly
\begin{equation}
    \sigma_y(\bar{\tau})^2 \approx \num{5E-27} + \SI{5E-27}{\second\squared} / \bar{\tau}^2
    \qc
\end{equation} 
where $\bar{\tau}$ is the averaging time in seconds, in the range \SIrange{0.1}{100}{\second} (again, see \cite{asmar:2012} where the \gls{psd} of phase noise at higher frequencies is also provided). Combining the two, the \gls{uso} noise can be approximated by
\begin{equation}
    S_{y_i^\epsilon}(f) \approx \qty(2\pi\times \SI{ e-14}{\per\hertz\tothe{0.5}})^2 \qty(\frac{f}{\si{\hertz}}+\qty(\frac{f}{\si{\hertz}})^{-1})
    \qc
\label{equ:USO_PSD}
\end{equation}
valid typically in the \SIrange{0.01}{100}{\hertz} region.

This model is compared to actual \gls{uso} noise measurements in a controlled environment obtained from CNES in \cref{fig:PHARAO}. The \gls{uso} shown is the flight model that will fly on the ISS in the ACES/PHARAO mission (launch expected in early 2024). It is coupled to a \SI{100}{\mega\hertz} VCXO to decrease the noise at high frequency (hence the PLL ``bump'' at around \SIrange{50}{100}{\hertz}). One notices the good agreement with the simple model in \cref{equ:USO_PSD} in the \SIrange{0.01}{e3}{\hertz} region. This agreement is somewhat surprising, given the different \gls{uso} types under consideration (NASA-Cassini/GRAIL vs. CNES-PHARAO) and given the poor information which the model is based on, but it's also reassuring as it demonstrates the good representativity of the model. Note that at low frequency ($< 0.01$~Hz) there is a discrepancy between the simple model \cref{equ:USO_PSD} and the PHARAO-USO measurements. It is not clear whether this is due to to an additional stochastic noise process, or a deterministic effect, but of no relevance for this work.

\appendix

\begin{acknowledgments}
The authors thank Marie-Christine Angonin, Adrien Bourgoin, Gerhard Heinzel, Christophe Le Poncin Lafitte, Martina Muratore and Kohei Yamamoto for useful discussions. We also thank François-Xavier Esnault from CNES for having provided the information on the ACES/PHARAO USO of \cref{sec:clock}.

O.H., A.H., M.L. and P.W. gratefully acknowledge support by Centre National d'\'Etudes Spatiales (CNES).

O.H. and M.S. gratefully acknowledge support by the Deutsches Zentrum für Luft- und Raumfahrt (DLR, German Space Agency) with funding from the Federal Ministry for Economic Affairs and Energy based on a resolution of the German Bundestag (Project Ref.~No.~50OQ1601 and 50OQ1801).

J.B.B. was supported by an appointment to the NASA Postdoctoral Program at the Jet Propulsion Laboratory, California Institute of Technology, administered by Universities Space Research Association under contract with NASA. Part of this research was carried out at the Jet Propulsion Laboratory, California Institute of Technology, under a contract with the National Aeronautics and Space Administration (80NM0018D0004).
\end{acknowledgments}

\bibliographystyle{apsrev4-1}
\bibliography{references}

\end{document}